\newif\ifpra
\prafalse

\def\dzero{D\O}
\def\mathunit#1{\mathop{\hbox{#1}}\mathclose{}\mathord{}}
\def\gev{\mathunit{GeV}}
\def\tev{\mathunit{TeV}}
\def\us{\mathunit{$\mu$s}}
\def\um{\mathunit{$\mu$m}}
\def\ns{\mathunit{ns}}
\def\cm{\mathunit{cm}}
\def\s{\mathunit{s}}
\def\xpb{\mathunit{pb}}
\def\xfb{\mathunit{fb}}
\def\ipb{\xpb^{-1}}
\def\ifb{\xfb^{-1}}
\def\gevc{\gev \kern -1.7pt/ \kern -1.7pt c}
\def\gevcc{\gevc^2}
\def\ra{\rightarrow}
\def\bbar{{{\bar b}}}

\def\tbar{{{\bar t}}}
\def\pt{{{p_{\scriptscriptstyle T}}}}
\def\et{{E_T}}
\def\mete{\setbox0=\hbox{$E$}%
          \hbox{$E$\rlap{\kern -0.5em\raise0.08\ht0\hbox{/}}}}
\def\met{{\mete_T}}

\ifpra
\documentstyle[times,pramana,psfig,floats]{ias}
\else
\documentclass{article}
\usepackage{psfig}
\fi
\begin{document}
\title{Prospects for Higgs Search at \dzero}
\ifpra
\author{Scott S. Snyder (for the \dzero\ Collaboration)}
\address{Brookhaven National Laboratory, Upton, NY, USA}
\else
\author{Scott S. Snyder (for the \dzero\ Collaboration)\\
Brookhaven National Laboratory, Upton, NY, USA}
\fi

\ifpra\else
\maketitle
\fi

\abstract {
The status of the Higgs search at the upgraded \dzero\ detector
is discussed.
}

\ifpra
\maketitle
\fi

\section{Introduction}

A major goal of experimental high-energy physics in the next decade
is to characterize the Higgs sector of the Standard Model.
The minimal version of the Standard Model requires a single scalar
particle, though many extensions predict additional particles.
Fits to the global set of electroweak data provide an indirect
upper limit on the mass of a scalar Higgs boson of
$m_H < 195\gevcc$, with a preference for low masses, with the minimum
at $m_H = 81\gevcc$~\cite{hbs,ewwg}.
Direct searches
at LEP have ruled out a scalar Higgs boson $m_H < 114.4\gevcc$.
These results provide a strong motivation to search for Higgs
bosons at the Tevatron.  This article summarizes the
state of the Higgs searches at \dzero\ as of summer 2002.

\section{The Run II \dzero\ Detector}

Both the Tevatron collider and the collider experiments have been
significantly upgraded since the last run.  The Tevatron was
augmented with a new injector, the collision energy was raised
from $1.8$ to $1.96\tev$, and the bunch spacing decreased from
$3.6\us$ to $396\ns$.
The goal for the current collider run is
to accumulate $10$--$15\ifb$ per experiment.
By the
end of 2002, the peak collider luminosity was over
$3\times 10^{31}\cm^{-2}\s^{-1}$, beating the record from Run I.
As of that date, \dzero\ had recorded about $50\ipb$ of data,
at a peak rate of about $1\ipb$ per day.  These rates are
expected to increase with further collider improvements.

Since Run I, the central tracker of the \dzero\ experiment has been
completely replaced.  It now consists of a silicon vertex detector
with four superlayers surrounded by a 16-layer scintillating fiber
tracker, using $835\um$ diameter fibers, the whole immersed in a 2T solenoidal
magnetic field.  Preshower detectors have also been added in front
of the calorimeters.  The calorimeters themselves are the same
as Run I, except that some of the electronics has been replaced
to handle the reduced bunch spacing.  The muon system has also been
significantly upgraded to improve coverage and to allow better
triggering in high luminosity environments.  The trigger and 
data acquisition systems were also replaced.  As of the end of 2002,
all detector components were operating well, except some
trigger systems which remain be commissioned.

\section{Higgs Production at the Tevatron}

A comprehensive study of the potential
for discovering Higgs at the Tevatron was carried out by the
Higgs Working Group (HWG)~\cite{hwg}
based on a parameterized
detector simulation that assumed averaged properties of the two detectors.
Unlike at the LHC, here there is no single channel in
which a significant result
can be expected.  Rather, we must combine results from all
decay modes, and also combine the data from the two experiments.
A $5\sigma$ discovery for $m_H=120\gevcc$ would then be
expected to require about $15\ifb$ of data.

The $gg\ra H$ mode has the largest cross section (about $1\xpb$
for $m_H=120\gevcc$).  However, for a light Higgs $m_H<140\gevcc$,
the dominant decay mode is $H\ra b\bbar$, rendering this mode hopeless
against the large backgrounds of $b\bbar$ production from other sources.
In this mass range, one must look for Higgs bosons produced in association
with a $W$ or $Z$ boson, the latter then decaying leptonically
(reducing the available cross section by about a factor of 30).
For higher mass Higgs bosons, the process $gg\ra H\ra WW$ may
be feasible.  Modes involving Higgs bosons produced with a heavy
quark pair ($b\bbar$ or $t\tbar$) have spectacular final states,
but very small cross sections ($< 10\xfb$).
However, some extensions to the Standard Model predict enhancements
to the $Hb\bbar$ and $H\ra \gamma\gamma$ modes.

\section{\dzero\ Status}

The \dzero\ experiment has repeated the analyses of the
HWG, using the full \dzero\ detector simulation.
The conclusions are roughly the same.  Work is in progress
on applying more sophisticated analysis techniques, such as
neural networks, to the problem.

For the $H\ra b\bbar$ channels, identifying $b$ jets is
crucial, both by using displaced vertices and by looking for
soft leptons from $b$-quark decay.  Monte Carlo (MC) simulations
show that displaced vertex tagging can have efficiencies
as high as $60\%$ for high-transverse energy ($\et$) jets, with a fake rate
from light quarks on the order of a few percent.  The tagging
performance depends on the track impact parameter resolution,
which has been measured in current data
to be $20\um$, already close to MC expectations.
The work of
determining the alignment of the tracking components
is still in progress, so this will improve.

For the $W(\ra e\nu/\mu\nu) H(\ra b\bbar)$ channels,
one looks for a final state containing
exactly one high-$\pt$ lepton, large missing $\et$ ($\met$), and
two $b$ jets.  The major physics background to these channels
is $Wb\bbar$; other significant backgrounds include $t\tbar$
and $WZ$.  An example of the results for the channel is
shown in figure~\ref{fg:mbb}.  Here, we require
$\et(e) > 20\gev$, $\met > 20\gev$, two tagged $b$-jets
with $\et>15\gev$, and for the third jet (if any),
$\et(j_3) < 25\gev$.  This last requirement rejects $t\tbar$
background.
The resulting $S/\sqrt{B}$ of 0.20 for the electron channel
is comparable to the
expectations from the HWG, but should be improved
significantly with the use of more sophisticated, multivariate
analysis methods.

\begin{figure}[htbp]
\centerline{\psfig{file=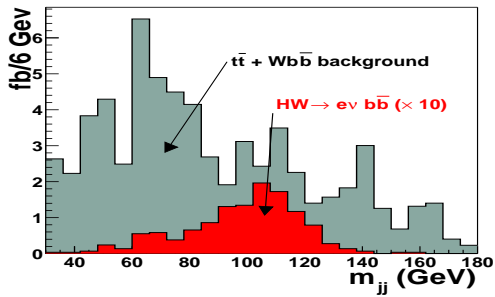,width=7cm,height=4cm}}
\caption{$m_{b\bbar}$ distribution from the $WH\ra e\nu b\bbar$ MC analysis,
using full detector simulation.  Plotted are the contributions
from the principal backgrounds ($Wb\bbar$ and $t\tbar$) and the
signal expectation for $m_H=120\gevcc$ (scaled up by a factor
of 10).}
\label{fg:mbb}
\end{figure}

For $Z(\ra ee/\mu\mu) H(\ra b\bbar)$,
one requires two oppositely
charged isolated leptons with a mass consistent
with a $Z$~boson and two $b$ jets.  A significance comparable
to the $WH$ channels is achievable though the overall event
rate is lower.  The $\nu\overline{\nu} b\bbar$ channel
is also powerful since the branching fraction of a $Z$~boson to neutrinos
is three times larger than that for any single lepton flavor.
For this channel, one requires the $b$ jets plus large $\met$.

The $H\ra WW^*\ra\ell\ell\nu\nu$ channels are
important for higher mass Higgses, $m_H > 130\gevcc$,
where the decay $H\ra WW$ is significant.  Important
backgrounds include $WW$, $t\tbar$, $W/Z$+jets, and QCD.
In our preliminary analysis, we require two electrons
with $\et>20\gev$, $m(ee) < 78\gevcc$, $\met > 20\gev$, and
no jets with $\et > 20\gev$.  In addition, since the $W$~boson
pair comes from the decay of a spin-0 particle, spin
correlation variables are useful, like the azimuthal angle between
the leptons ($\Delta\phi(\ell\ell)$)
(see figure~\ref{fg:ww}).

\begin{figure}[htbp]
\centerline{
\psfig{file=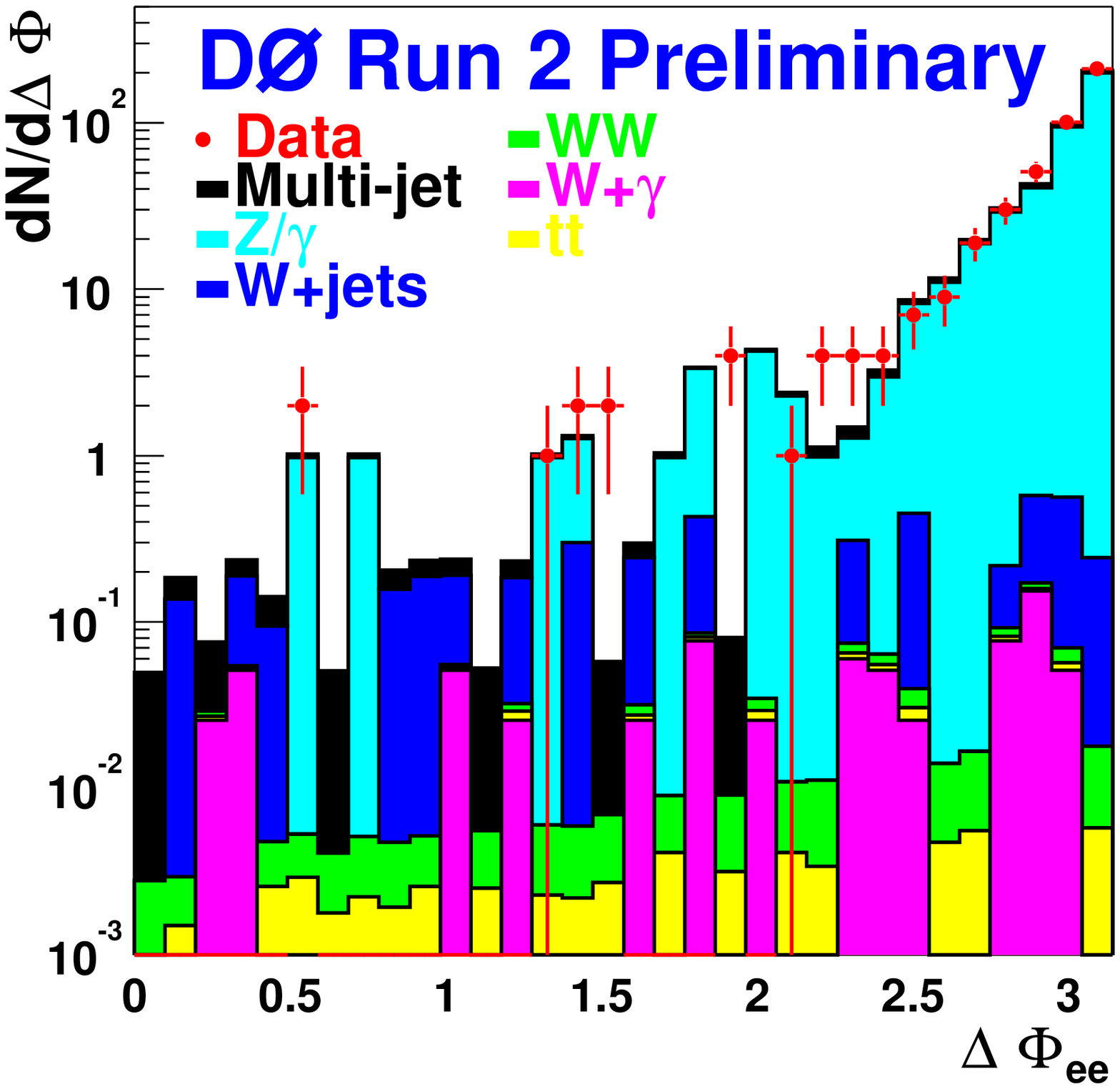,width=6cm,height=4cm}
\psfig{file=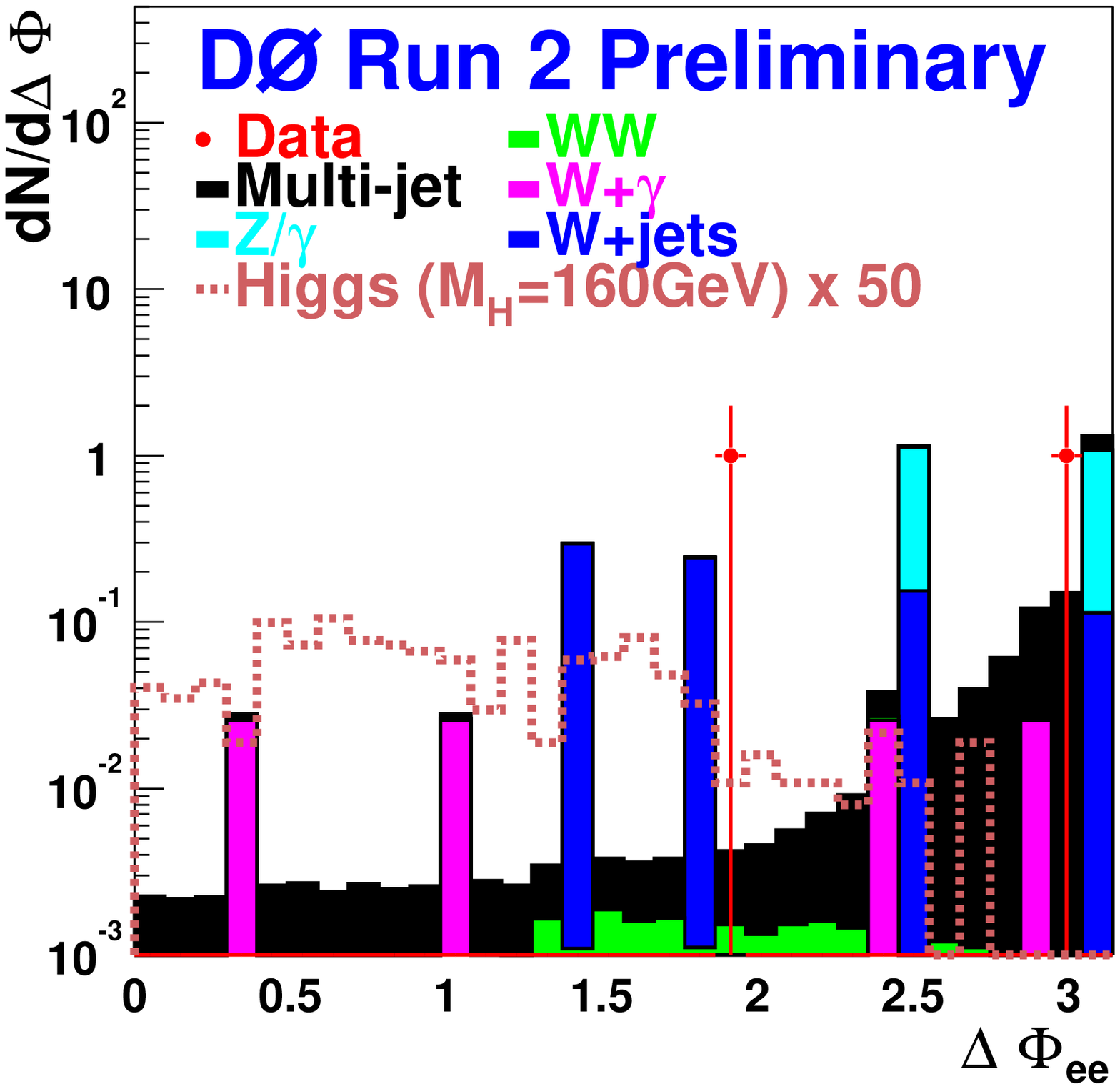,width=6cm,height=4cm}
}%
\caption{Results from the $H\ra WW^*\ra ee\nu\nu$ analysis,
plotting the $\Delta\phi(ee)$ distribution for the data and
the principal backgrounds.  The left plot is after the electron
$\et$ cuts only, the right plot is after all section cuts.
The right plot also shows the signal expected for $m_H=160\gevcc$
(scaled up by a factor of 50 to make it more visible).}
\label{fg:ww}
\end{figure}

\section{Summary}

The \dzero\ experiment is recording physics quality data.  Both the detector
and the accelerator performance are continually improving.
We are studying issues such as 
the $b\bbar$ mass resolution, $b$-jet tagging
efficiency, missing $\et$ resolution, and backgrounds to Higgs
processes.  We look forwards to seeing exciting
results!

\end{document}